\documentclass{PoS}
\usepackage{cancel}

\title{Interpretation of the top-quark mass results}

\ShortTitle{Interpretation of the top-quark mass results}

\author{\speaker{Gennaro Corcella}\\
  INFN, Laboratori Nazionali di Frascati\\
  Via E.~Fermi 40, 00044 Frascati (RM), Italy\\
        E-mail: \email{gennaro.corcella@lnf.infn.it}}

\abstract{I discuss recent work aimed at interpreting
  the top-quark mass measurements at the LHC and determining
the theoretical uncertainty.}

\FullConference{XXV International Workshop on Deep-Inelastic
  Scattering and Related Subjects\\
		3-7 April 2017\\
		University of Birmingham, UK}

\begin{document}
\section{Introduction}
The top-quark mass is one of most important quantities of the
Standard Model: it plays a role in the electroweak precision
tests \cite{gfitter}, since,
together with the $W$ mass, it
constrained the Higgs mass even before its discovery,
as well as in the property of the 
Standard Model vacuum to lie on the border
between stability and metastability regions \cite{degrassi}.
Such results, however, rely on the assumption 
that the top-quark mass world average, i.e.
$m_t=[173.34\pm 0.27{\rm (stat)} \pm 0.71{\rm (syst)}]$~GeV \cite{wave},
corresponds to the pole mass and no 
further uncertainty associated with the interpretation of the
measurements is accounted for.

In fact, the top-quark mass is determined
by comparing experimental data
with theory predictions: the extracted mass is the quantity 
$m_t$ in the calculation. 
In the following, I shall review the main methods used to reconstruct
the top-quark mass at the LHC and discuss the ongoing work to
interpret the top-mass results and determine the
theoretical uncertainty. I will finally make some concluding remarks.

\section{Interpreting the top-quark mass measurements}

\subsection{Standard measurements}

Top-quark mass extraction strategies are typically labelled as standard or
alternative measurements.
The standard ones, based on the template, matrix-element
and ideogram methods (see, e.g., the analyses in \cite{atlas1,cms1}),
compare final-state distributions, associated with
top-decay ($t\to bW$) products, with
Monte Carlo generators,
such as HERWIG \cite{herwig} or PYTHIA \cite{pythia}.
These programs simulate the hard scattering at leading order (LO),
multiple radiation in the soft or collinear
approximation and neglect the interference between
top-production and decay phases.
Most recent codes, such as 
aMC@NLO \cite{mcnlo} and POWHEG \cite{powheg}, 
implement the hard scattering at next-to-leading order (NLO), but still rely on
HERWIG and PYTHIA for showers and hadronization.

Much debate has been taking place through the years to connect
the measured mass in standard methods, often
called `Monte Carlo' mass, with theoretical definitions
like pole or $\overline{\rm MS}$ masses (see
\cite{corc} for a review).
On the one hand, unlike electrons, quarks do not exist
as free particles and this is reflected in the fact that
the pole mass exhibits the so-called renormalon
ambiguity \cite{beneke}, associated with the infrared divergences in the
higher-order corrections to the quark self energy.
The induced uncertainty on the pole mass has been
estimated to be roughly ${\cal O}(\Lambda_{\overline{\rm MS}})$. 
Such an ambiguity has been lately reconsidered, in view of the recent
calculation of the four-loop relation between pole and
$\overline{\rm MS}$ masses, the latest being free from renormalons
\cite{mspole}.
Refs.~\cite{nasben} and \cite{hpole} computed the ultimate uncertainty on
the top pole mass due to renormalons, obtaining 
about 110 and 250 MeV, respectively.
Ref.~\cite{hpole} 
estimated the uncertainty by analyzing order by order the pole-mass
series as a function of the $\overline{\rm MS}$ mass, while
Ref.~\cite{nasben} compared the exact result in \cite{mspole} with the
asymptotic expansion in \cite{beneke} and gauged the error
by relying on a prescription on the computation of an inverse Borel
transform.
Though differing by more than a factor of 2, nevertheless,
both evaluations are
smaller than the current uncertainty on $m_t$. 

On the other hand, since the
pole mass is defined in such a way to
reabsorb all radiative corrections, the invariant mass
of the products of the decay an on-shell top quark, as the top quark is
treated in standard Monte Carlo codes,
must be close to the pole mass.
Nevertheless, much work has been carried out to determine the
uncertainty on the measured mass and its identification with the top pole
mass. One of the main contributions to the theoretical error is
colour reconnection, namely the possibility that bottom quarks
in top decays
form a hadronic string (PYTHIA) or cluster (HERWIG) with (anti)quarks which
do not belong to their own showers, but rather to initial-state
radiation or $W$ decays. In the world-average analysis,
colour reconnection accounts
for about 300 MeV on the total error on the top mass.
Such a phenomenon clearly spoils
the association of the mass of the top-decay products with the pole mass and
has been investigated in \cite{spyros} in terms of the PYTHIA string model
and in \cite{corc1} by using the HERWIG cluster model.
In particular, Ref.~\cite{corc1} compared
standard $t\bar t$ events with those yielded by the simulation of
fictitious top-flavoured hadrons. In fact, if a
$T$-hadron decays according to the spectator model, the
$b$ quark is forced to form a cluster with either the spectator quark or
the products of its own shower, while the
colour connection with the initial state is suppressed.
The investigation in \cite{corc1}, currently in progress,
is also pretty interesting since, by applying lattice or Non Relativistic
QCD, one is able to connect the mass of a top-flavoured hadron with
any $m_t$ definition and it may thus serve as another benchmark
analysis to assess the uncertainty on the Monte-Carlo-driven
reconstructed top mass. 

Furthermore, work has been undertaken to confront Monte Carlo
distributions with resummed calculations performed in the
framework of Soft Collinear Effective Theory (SCET), using the so-called
MSR top mass definition, $m_{\rm MSR}(R)$, 
which, according to the value of $R$, may correspond, e.g.,
to the $\overline{\rm MS}$ or the pole mass.
Ref.~\cite{hoang} expressed the pole mass in terms of the 
SCET jet mass in $e^+e^-\to t\bar t$
collisions, assuming that $m_J(\mu)$
should mimic the reconstructed $m_t$ for $\mu\simeq Q_0$,
with $Q_0$ being the shower cutoff.
A shift about $\delta m\sim {\cal O}(\alpha_S\Gamma)\sim$~150-200~MeV,
where $\Gamma$ is the top width, was then determined.
More recently, Ref.~\cite{buten} compared PYTHIA with a 
SCET calculation at NLO, including the resummation
of next-to-next-to-leading soft- and collinear-enhanced logarithms
(NNLL), for the 2-jettiness in 
$e^+e^-\to t\bar t$ processes, trying to calibrate the MSR mass in
the resummation to reproduce the Monte Carlo spectrum.
In fact, it was obtained that, within the error range,
the mass parameter in PYTHIA is consistent with the tuned 
$m_{\rm MSR}(1~{\rm GeV})$, whereas it differs
by about $(0.57\pm 0.28)$~GeV from the corresponding pole mass.
The work in \cite{buten} was extended  
to $pp$ collisions in \cite{groom}, where
the extraction of
$m_t$ from boosted top jets with light soft-drop
grooming is proposed.
By comparing a NLL resummation
for the groomed top-jet mass 
with PYTHIA, 
the pole mass was found about 400-700 MeV below the calibrated
PYTHIA mass.
Another approach was suggested 
in \cite{moch1}: one measures an observable,
e.g. a total or differential $t\bar t$ cross section, without
any assumption on the Monte Carlo parametrization,
and, by comparing the data with the simulation,
calibrates both observable
and $m_t$.
The conclusion of Ref.~\cite{moch1} is that, with the current
precision on the inclusive $t\bar t$ cross section,
the uncertainty on this calibration is roughly 2 GeV.

Sticking to Monte Carlo programs,
a major improvement 
has been the recent release of
the $b\bar b4\ell$ generator \cite{bb4l}, which, within the POWHEG-BOX
framework,
simulates the full NLO process
$pp\to b\bar b\ell^+\nu_\ell\ell^-\bar\nu_\ell$,
including the interference between top production and decay,
and non-resonant contributions. It will be therefore very interesting
using this generator in template or matrix-element analyses and
comparing the results with those yielded by standalone
HERWIG or PYTHIA, as well as aMC@NLO or POWHEG with LO top decays.
Furthermore, the feasibility of POWHEG to be interfaced with both
HERWIG and PYTHIA should also shed light on the Monte Carlo systematics,
due to the use of different parton showers and hadronization models. 

\subsection{Alternative measurements}

Other strategies to measure $m_t$, making use of 
total or differential
cross sections, endpoints, energy peaks or kinematic
properties of $t\bar t$ final states,
are traditionally called `alternative' measurements.
The total $t\bar t$
cross section was calculated in the NNLO+NNLL approximation \cite{alex}
and allows a
direct determination of the pole mass.
Both ATLAS and CMS Collaborations have measured the
top mass from the inclusive cross section obtaining
$m_t=\left(172.9^{+2.5}_{-2.6}\right)$~GeV (ATLAS) \cite{sigmaatl} and
$m_t=\left(173.6^{+1.7}_{-1.8}\right)$~GeV (CMS) \cite{sigmacms},
  combining 7 and 8 TeV data.
In principle, even this extraction depends on the
Monte Carlo program used for the evaluation of the acceptance, 
but nonetheless the sensitivity to the implemented top mass turned out
to be very mild.
The errors in \cite{sigmaatl} and \cite{sigmacms}
are larger than those yielded by the standard methods;
however, they are expected to decrease thanks to the higher
statistics foreseen at the LHC Run II.
The NNLO calculation of the $t\bar t$ cross section has been
extended to differential distributions in \cite{mitov} and the
D0 Collaboration used it to measure the top-quark pole mass from
the $t\bar t$ invariant mass or transverse momenta spectra, finding
$m_t=(169.1\pm 2.5)$~GeV \cite{d0}.
The top pole mass was also determined from the
measurement of the $t\bar t+1$~jet cross
section, which is more sensitive to
$m_t$ than the inclusive $t\bar t$ rate,
following Ref.~\cite{ttj}, where the NLO $t\bar tj$ cross section was
calculated through the POWHEG-BOX and matched to PYTHIA.
The results are
$m_t=(173.70^{+2.28}_{-2.11})$~GeV (ATLAS) \cite{atlttj}
and $m_t=(169.90^{+4.52}_{-3.66})$~GeV (CMS) \cite{cmsttj};
the impact of the Monte Carlo input mass in the evaluation of the 
acceptance is negligible.
Lately, the NLO $t\bar t+1$~jet cross section has been calculated
in terms of the $\overline{\rm MS}$ mass and compared
with the LHC measurements \cite{fuster}.
The extracted value of
the $\overline{\rm MS}$ mass is nevertheless 
consistent with the value which can be obtained from the pole mass.

Other proposed methods to reconstruct $m_t$ 
rely on kinematic properties of
top-decay final states and hence, once again, the extracted mass must
be close to the pole mass. Unlike the invariant mass
of the top-decay products, used in the template analyses, quantities
like energy peaks, endpoints or purely leptonic observables
are however expected to exhibit a
larger uncertainty due to higher-order corrections, which are not anymore
reabsorbed in the mass definition.

In detail, it was found that the peak of the energy of the $b$-jet
in top decay at LO is independent of the boost 
from the top to the laboratory frame, as well as of the
production mechanism \cite{roberto}.
The CMS Collaboration measured the top mass from the $b$-jet
energy peak and obtained 
$m_t=\left[172.29\pm~1.17~({\rm stat.})\pm~2.66~({\rm syst.})\right]$~GeV
at 8 TeV \cite{bj}.

The $b$-jet+lepton invariant-mass
($m_{b\ell}$) spectrum was used by CMS to reconstruct 
$m_t$ in the dilepton channel: by comparing it with
PYTHIA, $m_t=(172.3\pm 1.3)$~GeV
was found \cite{mbl}.
The NLO calculation of $m_{bl}$ \cite{melnikov}, performed in the
narrow-width approximation
with the pole mass, is also available and 
exhibits some disagreement with respect to LO parton showers
\cite{corc1,mescia}.

The endpoints of distributions like 
$m_{b\ell}$, $\mu_{bb}$ and $\mu_{\ell\ell}$, 
where $\mu_{bb}$ and $\mu_{\ell\ell}$
are related to the $b\bar b$ and $\ell^+\ell^-$ invariant masses
in the dilepton channel,
were also explored to constrain $m_t$ \cite{end}.
Since $b$-flavoured jets can be calibrated directly
from data, Monte Carlo uncertainties on the endpoints
are mostly due to colour reconnection. 
The result, based on LO kinematics, is
$m_t=\left[173.9\pm 0.9 ({\rm stat.})^{+1.7}_{-2.1}({\rm syst.})\right]$~GeV.
Updating the $B$-energy peak, $m_{b\ell}$ and endpoint 
analyses using novel data as well as the late NLO generator $b\bar b4\ell$
\cite{bb4l}
is certainly worthwhile to be pursued for the sake of a more reliable
estimate of the theoretical uncertainty.

Finally, purely leptonic observables in the dilepton channel,
such as the Mellin moments of lepton energies or transverse
momenta, 
were proposed to measure $m_t$ as they do not require the
reconstruction of the top quark \cite{frix}.
Such quantities do exhibit small
hadronization effects, but they are sensitive to the
production mechanism, to the Lorentz boost from the top rest frame
to the laboratory frame, as well as 
to higher-order corrections.
Preliminary analyses have yielded $m_t=\left[171.70
  \pm 1.10 ({\rm stat.})^{+2.68}_{-3.09}({\rm syst.})\right]$~GeV \cite{cmslep}
(CMS, based on LO MadGraph)
and $m_t=(173.2\pm 1.6)$~GeV \cite{nisius} (ATLAS, based on the MCFM NLO code
\cite{mcfm}) and are expected to be improved by
matching NLO top-decay amplitudes with shower/hadronization
generators.

\section{Conclusions}

I discussed the interpretation of the top mass measurements
at the LHC: template-based
determinations, relying on the reconstruction of the invariant mass
of the
top-decay products, yield results close to the top-quark pole mass,
but nevertheless a careful determination of the theoretical uncertainty,
of both perturbative and non-perturbative origin, such as
missing higher orders, width corrections and
colour-reconnection effects, is compelling.
The late implementation of top decays and interference between
production and decay phases at NLO should help
to quantify the perturbative error on $m_t$.
As for non-perturbative corrections, studies on colour
reconnection or simulations of final states 
where the tops are forced to hadronize and the bottom quarks
to form colour-singlet clusters with final-state partons
will be useful to address the hadronization
systematics.

As for top-mass definitions, two papers have addressed
the renormalon ambiguity on the pole mass: although they
disagree by roughly a factor of 2, the estimated uncertainties
are both below the current error on $m_t$.
Studies aimed at relating the extracted mass to the pole mass
have been carried out within the
SCET formalism for $e^+e^-$ processes and lately extended
to $pp$ collisions.
Alternative measurements, based on the comparison of the
$t\bar t$ and $t\bar t j$ cross sections with
NLO or NNLO calculations,
allow a clean extraction of the pole or $\overline{\rm MS}$ mass,
with errors which will decrease once the LHC statistics get higher.
Other strategies, relying on kinematic properties of top-decay final
states, have so far employed parton-shower generators with LO top decays:
updates using NLO codes will lead to a
a more reliable estimate of the theoretical uncertainty.

In summary, given the latest LHC performances, top-quark phenomenology
is on the road to become precision physics and the measurement
of the top mass to reach a very high level of accuracy.
In view of the implementation of advanced event generators for top physics,
as well as of refined calculations for top production
and decay, more accurate determinations of the top mass and, in particular,
of the theoretical uncertainty are therefore both feasible and
desirable.

\end{document}